\documentstyle[aps,epsfig,graphics]{revtex}
\begin{document}

\title{Chaos in  Pseudo-Newtonian Black Holes with  Halos.}
\author{E. Gu\'eron\thanks{%
e-mail: gueron@ime.unicamp.br} \and P. S. Letelier\thanks{%
e-mail: letelier@ime.unicamp.br} }


\address{Departamento de Matem\'atica Aplicada, Instituto de
  Matem\'atica,
Estat\'{\i}stica e Computa\c{c}\~ao Cient\'{\i}fica, Universidade 
Estadual de
Campinas, 13083-970, Campinas, S\~ao Paulo, Brazil }

\maketitle

\begin{abstract}
 The Newtonian 
as well as the special relativistic dynamics are used to study the stability of
 orbits of a test particle moving around  a black hole plus a dipolar halo.
 The black hole is modeled by either the usual monopole potential or the
 Paczy\'nki-Wiita pseudo-Newtonian potential. The full general relativistic 
 similar  case is also considered.  The  Poincar\'{e} section method 
 and the Lyapunov characteristic exponents show  
 that the orbits for  the pseudo-Newtonian potential models are
 more unstable than the  corresponding  general relativistic geodesics.
  

\end{abstract}

\section{\bf Introduction.}
To consider  relativistic effects in many body simulations 
is not a simple task  due to the fact   that the metric 
representing their gravitational interaction is far from being 
known. For the simplest case of  two gravitating
 bodies the metric is known  numerically only for 
 few initial conditions and for a
limited amount of time [see for instance, Marronetti et al. (\cite{marro})].
Also, assuming  that the  metric is known, the use of the
 geodesic equations to determine the trajectory of the bodies
  represents a quite non trivial problem.

 In general, we have three main ways to consider 
complex systems:  a) A full numeric approach with its inherent
limitations due to the use of floating point arithmetics and
arbitrariness of discretizations of fundamentally continuous
functions and variables. Also we have rather unphysical ad hoc 
assumptions like the introduction of numerical viscosity.
b) The use of perturbative methods that are usually employed together
with drastic approximations like the mean field approximation for the 
potentials in many body simulations. These approximations introduce
irreversibility in an intrinsic reversible situation.
c) The modeling of the problem with simpler equations in which one takes into
 account  a few essential features of the problem. In general,
this model can be solved in a more exact form of the two
 precedent case.
But, we have changed  the initial problem for a simpler one that may
falsify  results. In other words, there is not a perfect method 
to solve a   a complex problem. We believe that all of them are valid
when the adequate cautions are taken. Furthermore, they
 are complementary
and the usually not proven mathematical or internal consistence of 
the methods can be independently checked at least for some particular cases.

Due to the weakness of the gravitational field, far from the
 particles' horizon, the Newtonian gravity is proven to be a 
reliable description of the gravitational interaction. 
One can simulate relativistic effects within the Newtonian theory 
changing the usual potentials  to take  into account the existence of the
horizon. In other words to model  relativistic effects via a
 pseudo-Newtonian potential.   These models are simpler enough to
describe complex systems that are far beyond of  todays 
knowledge of the full  general relativity, e.g.,  the  n-body 
simulation of the  collision of two galaxies to any
 degree of resolution. 

One of the simplest pseudo-Newtonian potentials  to
describe  behavior of test particles moving close to a black hole
is the  Paczy\'nski and Wiita (\cite{pw})  pseudo-potential,
\begin{equation}
\Phi =-\frac{GM}{R-R_{g}}.  \label{pw1}
\end{equation}
The addition of the term $R_{g}=2GM/c^{2}$ critically changes the
particles' trajectory near the source. Some results like the last stable
 circular orbit are
predicted in this model. Other pseudo-Newtonian models can be found in
literature, e.g., the one studied by Semer\'{a}k and Karas (\cite{sk}) to
describe rotating black holes, i.e., to approximate the Kerr solution.

We believe that the study of the  Paczy\'nski and Wiita (PW) 
 potential in simple albeit 
nontrivial situations may shed some light into the correctness of
the pseudo-potential approach. In particular, in this      article,  we 
 study integrability and chaos in a system that represents
a spherically symmetric source (monopole) surrounded by a dipolar halo
(external dipole), that is the simplest mean potential used to describe
astrophysical systems restricted to a core and halo, see for instance 
Binney and Tremaine (\cite{binney}).
 Different theoretical approaches are used to study this configuration.
 First we use Newton second law to find the motion equations for  test
particles ($\vec{a}=-\nabla \Phi)$ for  two different potentials
 that describe a core plus a dipolar halo system:
a) The standard monopole plus external dipole expansion that solves 
 the usual  Laplace equation that
 is totally integrable, see for instance 
Grammaticos et al.(\cite{dnewt}), and b) We
  replace in the former case
the monopole term by the  PW potential (\ref{pw1}). In this case  the 
trajectories are chaotic like in the
equivalent full general relativistic system, Vieira and Letelier
 (\cite{dipolo}).

We also analyze the equivalent cases using the special relativistic dynamics.
 We solve the equation 
 $a^{\mu }=F^{\mu }$ with 
 $a^{\mu }=\frac{d^{2}x^{\mu }}{d\tau ^{2}}=\gamma
 \frac{d}{dt}\left( \gamma\frac{dx^{\mu }}{dt}\right) $  and 
$F^{\mu }=\gamma
(-{\bf \nabla }\Phi \cdot {\bf v}/c,-{\bf \nabla }\Phi )$,
where $\gamma =(1-{\bf v}^{2}/c^{2})^{-1/2}$,
and $\Phi$ is taken   as in the Newtonian cases. We
 first use the
monopole plus dipole potential that solves the Laplace equation.  
A phase space analysis shows that the system is stable.  Replacing
the monopole term by the PW potential we obtain a very unstable system.
We also review the equivalent system in general relativity. The geodesic
equations for Schwarzschild monopole plus dipolar halo 
give us  chaotic trajectories in the phase
space as shown in Vieira and Letelier (\cite{dipolo}).

 In  each one of the studied cases  we have an  integrable
  Hamiltonian 
system  of equations for the motion of a test particle moving in a
spherically symmetric  attraction center 
 (standard monopole, PW potential or Schwarzschild metric) that is perturbed by an external 
 dipole term. In all these situations we
can apply the KAM (Kolmogorov, Arnold and Moser) theory, see for instance
Tabor (\cite{tabor}). 
Since our mass distribution has axial symmetry  we are
restricted to an effective 
 two-dimensional problem. In the integrable case, 
 in phase space,  the orbits of test particles  
  will be confined to a 2-torus. For a constant value of one
   of the coordinates  we obtain a
planar section of the  phase space. In the integrable case we shall see
closed curves for each initial condition, intersections of invariant tori.
 While in the  non-integrable case some tori will be destroyed and the 
 region will be  ergodically fulfilled.  In order to evaluate the 
degree of instability of the orbits in  each system we also compute
 the Lyapunov exponents that indicate us how initially close trajectories
separate.
                  
\medskip

\section{Newtonian dynamics.}

\medskip

The standard monopole plus external dipole potential in the 
usual cylindrical coordinates $%
(r,z,\phi )$ is 
\begin{equation}
\Phi =-\frac{GM}{\sqrt{r^{2}+z^{2}}}+D\;z,  \label{newt}
\end{equation}
where $D$ is the dipolar strength, $G$  the Newton constant, 
and $M$ the  mass of the attraction center. We use units such that 
$GM=1$, furthermore  we shall take $c=1$. From
 the angular momentum and energy conservation we find that  the
motion is restricted to the region defined by 
\begin{equation}
E^{2}-1 -\frac{L^{2}}{r^{2}} -2\Phi\geq 0.
\label{newtbound}
\end{equation}
 $L$ is the specific angular momentum of the test
 particle  and $E=\sqrt{1+2E_{mech}},$
where 
$$
E_{mech}=\frac{\dot{r}^{2}+\dot{z}^{2}}{2}+\Phi (r,z)+\frac{L^{2}}{2r^{2}}
$$ 
is the specific  energy. 
Note that $E$ become imaginary for $E_{mech}<-0.5$
that is the energy of a particle standing on the black hole horizon.
The phase space orbits are studied using the 
Poincar\'{e} section method. In Fig.1 we present the
 surface of section $z=0$
for the  constants: $L=3.9$,  $E=0.976$,  and $D=2\times 10^{-4}.$ 
This  surface section 
characterizes an integrable system as expected.

Now we shall replace the monopolar term by the
 PW pseudo-Newtonian potential, i.e.,
\begin{equation}
\Phi =-\frac{1}{\sqrt{r^{2}+z^{2}}-2}+D\;z.  \label{pwd}
\end{equation}
 Again, the motion of test particles 
 will be restricted to the region that solves (\ref{newtbound}) with $\Phi$
 given by (\ref{pwd}).
In Fig. 2 we present the surface of section  $z=0$. We take
 the  values for the constants as in the preceding case:  $L=3.9$,  $E=0.976$, 
 and $%
D=2\times 10^{-4}$.  Contrary to the previous case 
we  observe chaotic orbits in
this Poincar\'{e} section.

\section{\bf Special relativistic dynamics.}

In principle, the use of the special relativistic dynamics should improve
the modeling of general relativity with pseudo-Newtonian potentials,
see  Abramowicz et al. (\cite{abra}).       Although, these authors found that
the predicted spectra often differ rather substantially from those 
obtained in  the full general relativity context.
>From the relativistic motion equation we get
 \begin{eqnarray}
\frac{d}{dt}(\gamma +\Phi ) &=&0\Rightarrow \gamma +\Phi =E,  \label{energy}
\\
\frac{d\theta }{dt} &=&\frac{L}{\gamma \,r^{2}}.  \label{angular}
\end{eqnarray}
By using the above equations and  $u^\mu u_\mu =1$
  we obtain
\[
(E-\Phi )^{2}(1-\dot{r}^{2}-\dot{z}^{2}-\frac{L^{2}}{[(E-\Phi )r]^{2}})=1
\]
which is used to calculate the region in which the motion is confined.
Finally, the motion equations for the variables $r$ and $z$ are,
\begin{eqnarray}
(\Phi -E)\frac{d^{2}r}{dt^{2}} &=&\frac{\partial \Phi }{\partial r}(1-\frac{%
dr}{dt}^{2})-\frac{\partial \Phi }{\partial z}\frac{dz}{dt}\frac{dr}{dt}-%
\frac{L^{2}}{(E-\Phi )r^{3}},  \label{req} \\
(\Phi -E)\frac{d^{2}z}{dt^{2}} &=&\frac{\partial \Phi }{\partial z}(1-\frac{%
dz}{dt}^{2})-\frac{\partial \Phi }{\partial r}\frac{dz}{dt}\frac{dr}{dt}.
\label{zeq}
\end{eqnarray}
As in the previous section, we start with the usual monopole plus
external dipole potential field, i.e., we identify $\Phi $ with (\ref{newt}%
). In Fig. 3 we
 draw
the Poincar\'{e} section defined by the plane $z=0$. The constants
 are the
same of the  preceding section, $L=3.9$,  $E=0.976$, and  $D=2\times 10^{-4}$.
 We  notice that the tori were preserved in this case, we have stability of
  orbits. This is and indication of integrability of the system.

Now we start the study of the PW potential plus dipolar halo, i.e., 
we identify $\Phi $
with (\ref{pwd}). Unfortunately we cannot confine the orbits by using the
constants attributed to all the preceding cases. We put, $L=4.2$, $E=0.972$,
 and $D=4.2\times 10^{-4}$. Now the  Poincar\'{e}
section is taken as  $z=-5$. The figure in this case, Fig. 4, 
represents a very chaotic system. We used the same constants to draw another
Poincar\'{e} section for PW potential plus dipolar halo using  Newtonian
 dynamics. The results are presented in Fig. 5. We see some stable
islands in the negative $p_{r}$ region that cannot be observed Fig. 4.
We conclude then that the orbits obtained in the special relativistic context
are less stable than the ones obtained with Newton law.
The conjugated variables used  were $dr/dt$ and $r$. We made some tests
using $dr/d\tau $ and $r$.  The results were qualitatively the same.

\section{\bf General relativistic dynamics.}

We start from the axisymmetric line element
\begin{eqnarray}
ds^{2} &=&e^{2\psi (u,v)}dt^{2}-e^{-2\psi
(u,v)}(u^{2}-1)(1-v^{2})d\phi ^{2}  \label{weylxy} \\
 &&-e^{2(\gamma (u,v)-\psi (u,v))}(u^{2}-v^{2})\left( 
\frac{du^{2}}{u^{2}-1}+\frac{dv^{2}}{1-v^{2}}\right), \nonumber
\end{eqnarray}
in prolate  coordinates $(t,u,v,\phi)$.     The coordinates $u$ and $v$
are related to the usual cylindrical coordinated by $u=(R_++R_-)/(2m)$ and
 $v=(R_+-R_-)/(2m)$, where $R_\pm=[r^2+(z\pm m)^2]^{1/2}$ and m$=GM/c^2$.  The 
Schwarzschild monopole
plus   a dipolar halo is represented by
\begin{equation}
\psi (u,v)=\frac{1}{2}\log \left( \frac{1+u}{1-u}\right) +D\,uv.
\label{psisch}
\end{equation}
     Note that taken the limit, $\lim_{c^{-2}=0} \psi/c^{-2}$, with aid of
 l'H\^opital rule, we recover (\ref{newt}). To have the right units 
to take the limit we need to add a $c^{-2}$ factor to $D$.

The Einstein equations for these class of solutions as well as
 the corresponding
geodesic equations are studied in great detail in Vieira  and  Letelier
( \cite{vlastr}).
Due to the  axial symmetry of the metric again the effective geodesic dynamics
 of the test  particles is restricted to a three dimensional ``phase space''.

The Poincar\'{e} section is draw for $v=0$ (that is equivalent to $z=0$).
In Fig. 6 we present the section for the values of the constants  $L=3.9$,
 $E=0.976$, and $ D=2\times 10^{-4}$. Chaotic orbits may be observed 
 for instance
in the region indicated with a rectangle. A zoom of this region is presented 
 in Fig. 7. We can compare Fig. 6 with Fig. 2 and conclude that the
orbits obtained via geodesic equation in general relativity are more stable
than the ones obtained from the PW potential plus  dipolar halo in Newtonian
 and special relativistic dynamics.

\section{Lyapunov Exponent}

We shall study  the Lyapunov exponents for the systems above described  to
better analyze  the orbits stability. We shall use the Lyapunov
characteristic number (LCN) that is defined as the double limit 
 \begin{equation}\label{lcn}
LCN=\lim_
{\scriptsize
\begin{array}{l} \delta_0\rightarrow 0 \\
 t\rightarrow \infty 
 \end{array}
}
 \left[\log(\delta /\delta _{0}) \over t\right] , 
\end{equation}
 where $\delta _{0}$ and  $\delta $\ are the deviation of
two nearby orbits at times $0$ \ and $t$\, respectively. We get
the largest LCN using the technique suggested by Benettin et al. (\cite{bgs})
       
 We start comparing the LCN for orbits in a  PW+Dipole system 
in special relativity  and the LCN for orbits in a  PW+Dipole in
 Newtonian theory. The Constants are $L=4.1$,
 $E=0.972$, and $ D=4.1\times 10^{-4}$. The maximum LCN was obtained around
 $r=20,$ $ z=-5,$ and  $ p_r=-0.04$. Note that the  value of 
$p_z$ is determined by the
constants of motion and the value of $r,z$ and $p_r$. For the 
relativistic case we get
 $LCN=(3.2\pm 0.4)\times10^{-4}$  while for the Newtonian approach we obtain
 $LCN=(1.8\pm 0.4)\times10^{-4}$. We did some tests for the usual integrable
Newtonian  monopole plus  dipole system and we  always obtain for
 the LCN at least one  order of magnitude lower that the precedent  cases.

For orbits of test particles  in the the full  general relativistic
 monopole plus dipole system and in the Newtonian PW+Dipole system
 we  chose $L=3.902$, $E=0.9756$ and $ D=2.0\times
10^{-4}$. We obtain for orbits in the PW+Dipole system $LCN=(2.0\pm 0.5)\times10^{-4}$. 
This value was obtained for
orbits around $r=7.5, z=0,$ and $p_r=0$. For the general
 relativistic system the
proper time and the coordinate time were tested in the 
equation (\ref{lcn}) and no
significant difference was found. The largest $LCN$ was computed around
$u=9.75,$ $ v=0,$ and  $p_u=-0.038$.  As before,  $p_v$ is fixed by the
value of the other variables and the motion constants.  We  found always
  $LCN < 5\times10^5$. The Lyapunov like coefficients
used in general relativistic systems may have different forms as the one
suggested by Burd and Tavakol (\cite{bt}) in the study of Bianchi IX 
systems. However, we have studied a simple system with no
 singularities besides the black hole where we have a well
 defined evolution parameter. Hence,  in this case 
no significant difference should be found by using other definition  
of the Lyapunov coefficients. Furthermore, in the general relativistic system 
studied we have several natural ways to choose the 
space variables e.g.,  the spheroidal 
$(u,v,\psi)$ and  the cylindrical $(r,z,\phi)$. We found no significant
differences when either  system of coordinates are used to describe the
orbits of particles moving a few Schwarzschild radii apart from the central
black hole.

In summary, the study of Lyapunov coefficients confirms
 the qualitative analysis of the Poincar\'e section method,  we
 have that the general relativistic orbits
are more stable than the Newtonian and special relativistic
 ones. The special relativistic orbits are the most unstable.

\section{\bf Discussion}

In the Paczy\'nski-Witta potential, the term $-2GM/c^2$ in the denominator of
the equation (\ref{pwd}) creates a saddle point in the
 effective potential
 in Newtonian as well as in special relativistic  dynamics. The
addition  of the dipole term separates the stable and unstable
 manifold emanating
from the hyperbolic fix point as discussed by Letelier and Vieira (\cite
{periodic}). In this case, as consequence of the
Poincar\'{e}-Birkhoff theorem, 
there is an homoclinic web that
gives rise to  chaotic motion for bounded 
orbits in phase space, see for instance Tabor  (\cite{tabor}).

The chaotic orbits encountered in the pseudo-Newtonian plus dipole system
agrees with the general relativistic equivalent situation.
  However, those effects
might be distorted in the PW approach because the Poincar\'{e} sections
as   well as the Lyapunov exponents show
 more unstable orbits. This instability is magnified  when the special
  relativistic dynamics is used. 
 Vokrouhlick\'y and Karas \cite{vk} studied the stability of orbits
for particles gravitating around a $1/R$
  Newtonian potential with an axisymmetric perturbation.    Sridhar and
  Touma (\cite{st}) found for the same class of potentials 
that the instability decreases in orbits closer to
  the black hole. This result may
  not be verified when pseudo Newtonian or full general relativistic model are
  considered. The main difference being the presence of a saddle point
in the  effective potential near  the black hole. Therefore orbits near the
  core may be more unstable because of this critical point in the effective
potential that is a source of instability.

 \begin{acknowledgements}
The authors thank to FAPESP for financial support. EG thanks 
A.E. Motter  for  discussions about Poincar\'{e}-Birkhoff
theorem.
\end{acknowledgements}

\begin{twocolumn}
\begin{figure}
\epsfig{width=3in,height=2in,file=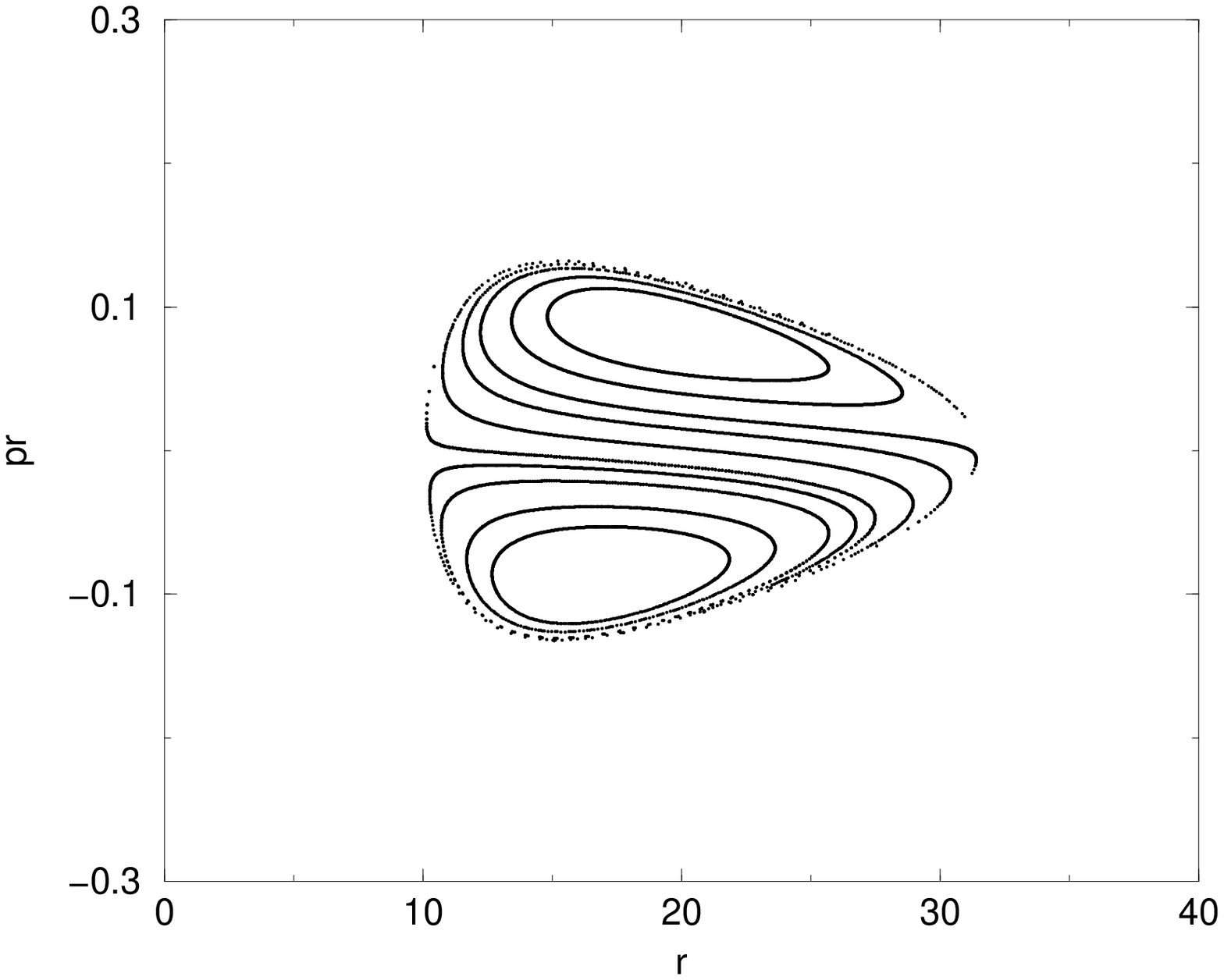}
 \caption{
  Surface of section for the Newtonian motion of a test particle in a standard
   monopole plus external dipole potential
   for $L_z=3.9$, $E=0.976$, and $D=2\times 10^{-4}$.  The section
   corresponds to the plane $z=0$.  For these values 
  of the parameters we have
 the section of an integrable motion.}

\end{figure}
\begin{figure}
\epsfig{width=3in,height=2in,file=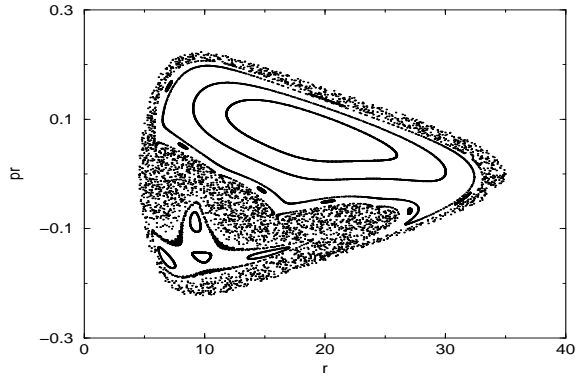} 
\caption{Surface of section for the Newtonian motion of a test particle in a
Paczy\'nski-Wiita potential 
 plus a dipolar halo for $L_z=3.9,$ $E=0.976$, and  $D=2\times 10^{-4}$.  The
  section corresponds to the plane $z=0$. We see chaotic motion. }
\end{figure}

\begin{figure}
\epsfig{width=3in,height=2in,file=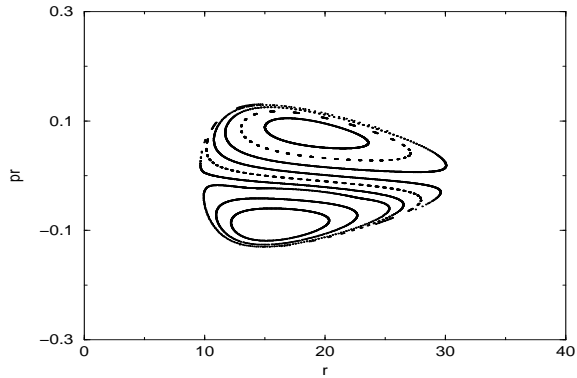} 
\caption{Surface of section for  the special relativistic  motion of a test
 particle in a usual monopole potential 
 plus a dipolar halo for $L_z=3.9$, $E=0.976$, and  $D=2\times 10^{-4}$.
  The section
  corresponds to the plane $z=0$. For these values 
  of the parameters we have
 the section of a regular motion}
\end{figure}

\begin{figure}
\epsfig{width=3in,height=2in,file=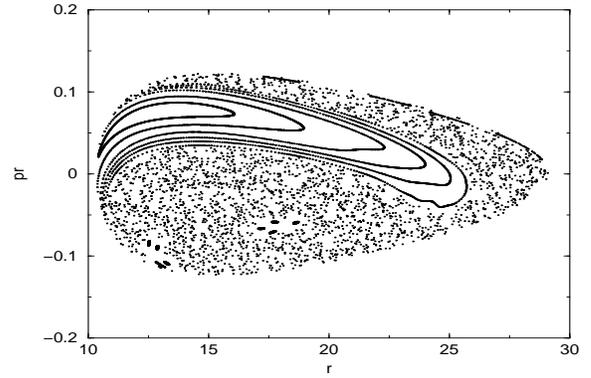} 
\caption{ Surface of section for the special relativistic  motion of a test
 particle in a
Paczy\'nski-Wiita potential 
 plus a dipolar halo for  
$L_z=4.2$, $E=0.972$, and  $D=4.2\times 10^{-4}$.  The section corresponds
 to the
 plane $z=-5$. We have a very irregular motion.}
\end{figure}

\begin{figure}
\epsfig{width=3in,height=2in,file=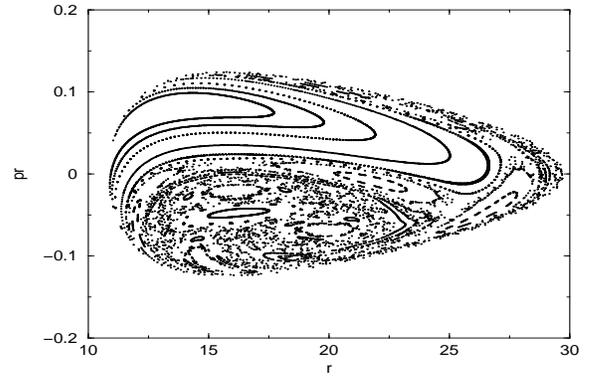} 
\caption{Surface of section of the Newtonian motion of a test particle in a
Paczy\'nski Wiita potential 
 plus a dipolar halo for  $L_z=4.2$, $E=0.972$, and  $D=4.2\times 10^{-4}$.
   The section corresponds to the plane $z=-5$. We have an irregular motion but
    it is more stable than the one shown in the precedent figure.}
\end{figure}

\begin{figure}
\epsfig{width=3in,height=2in,file=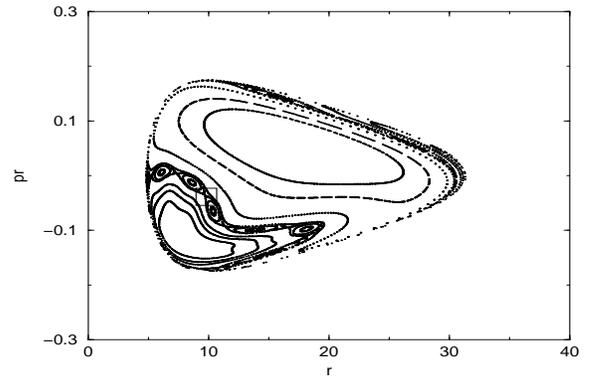}
 \caption{Surface of section for the geodesic motion of a test particle
 in a Schwarzschild monopole
   with  a dipolar halo for $L_z=3.9$, $E=0.976$, and  $D=2\times 10^{-4}$.
    The section corresponds to the plane $v=0$. For these parameters we have
     small regions of instability.}
\end{figure}

\begin{figure}
\epsfig{width=3in,height=2in,file=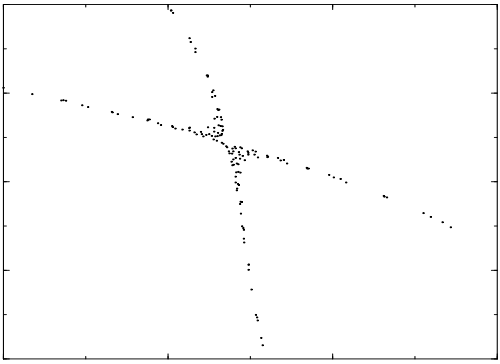}
 \caption{ A zoom of the small boxed region of the previous figure.}
\end{figure}
\end{twocolumn}

\end{document}